# Detailed dark matter maps of galaxy cluster substructure and direct comparison to simulations

Dan Coe
JPL/Caltech
coe@jpl.nasa.gov
(337) 281-1433

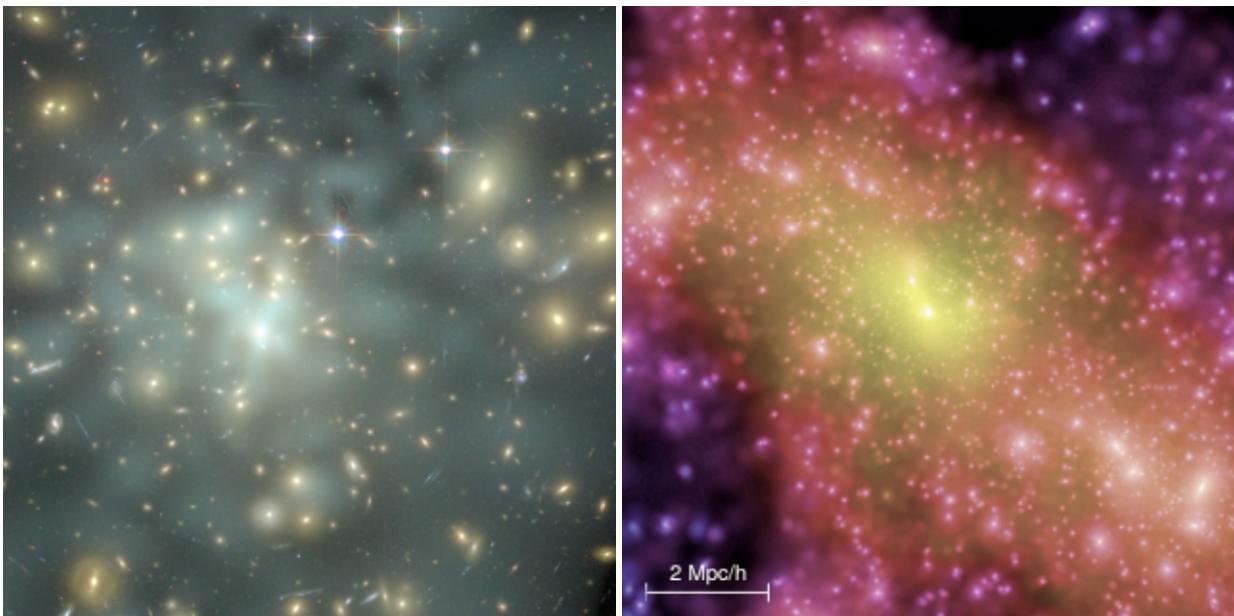

Observed dark matter mass map of Abell 1689    Simulated dark matter cluster from Millennium

White Paper Submission
to the
Cosmology and Fundamental Physics Panel
of the
National Academy of Sciences' Committee on Astro2010:
The Decadal Survey on Astronomy and Astrophysics

# Executive Summary

Images from the next generation of telescopes will enable strikingly detailed reconstruction of the dark matter distributions in galaxy cluster cores using strong gravitational lensing analysis. This will provide a key test of $\Lambda$CDM cosmology on cluster scales where tensions currently exist. Observed dark matter distributions will be compared directly to those realized in simulations, forgoing any assumptions about light tracing mass. The required observations are deep, multicolor, and high-resolution, ideally supplemented with spectra of faint objects. ACS onboard HST is capable of obtaining images of sufficient quality, but for prohibitive integration times. The next generation of telescopes promises to efficiently yield the required images. An analysis method capable of process the expected large numbers of multiple images has been developed (see below). The full range of constraints possible from analyzing these detailed mass maps is a matter of ongoing investigation.

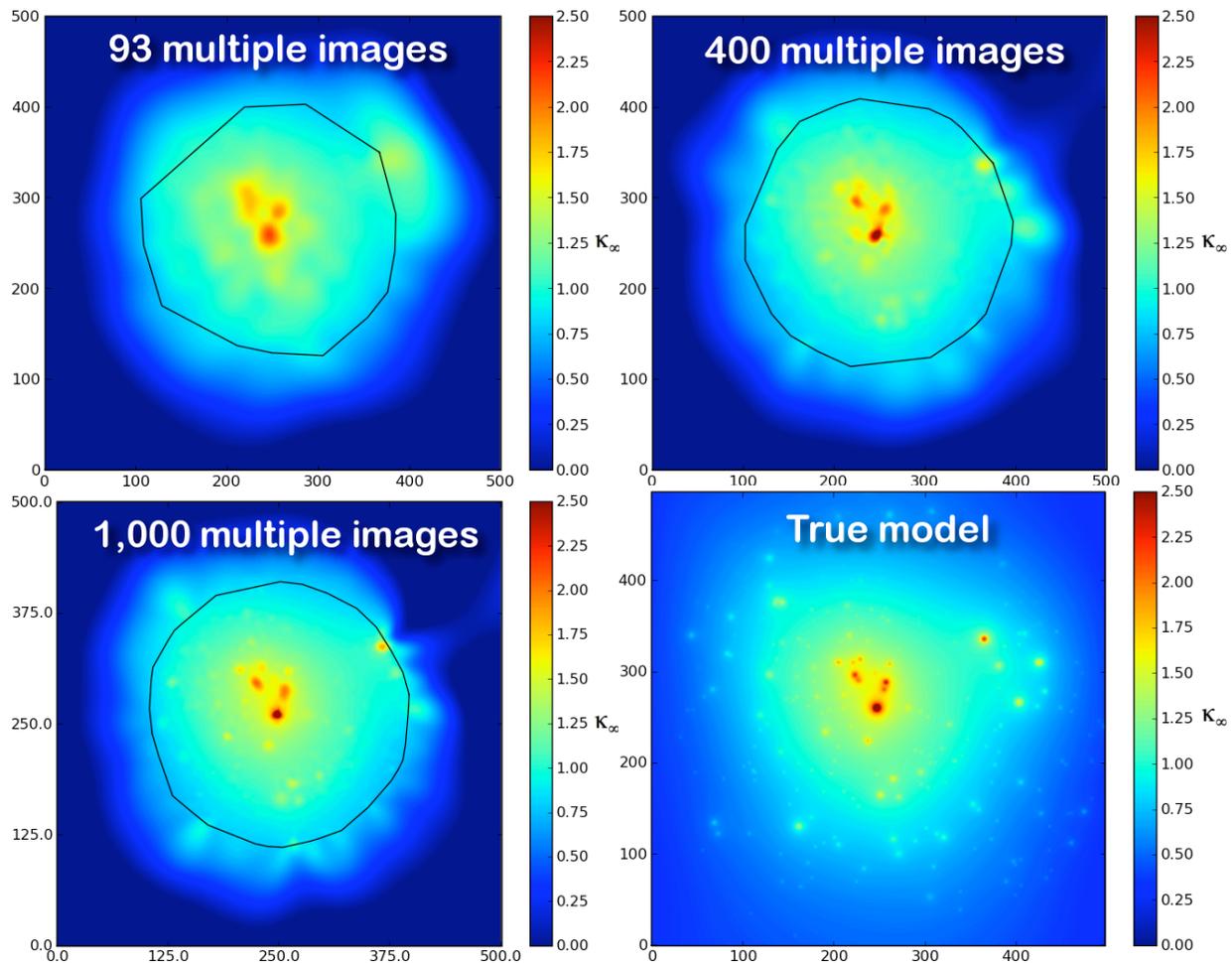

*Dark matter mass map recovery possible given different numbers of multiple images: 93, 400, and 1,000. Fine details are resolved in the latter cases. Recovery is poor outside the regions where multiple images are detected (black lines). The true mass map is shown at bottom right.*

# Introduction

As impressively as ΛCDM simulations appear to reproduce the large scale structure of the universe (Spergel et al. 2006), comparisons on smaller scales have so far proven less convincing. Compared to simulated galaxies, true galaxies appear to have "missing satellites" (Strigari et al. 2007, though Diemand et al. 2005 provide a mechanism for some subhalos to stay dark) or perhaps too much substructure (from our too-frequent observations of "flux anomalies" in gravitational lenses: Macciò & Miranda 2006, Diemand et al. 2007). A "cusp-core" controversy has also raged for years, though cusps appear to be evaporating in the latest simulations (Navarro et al. 2008).

On cluster scales we have yet to conclusively confirm or rule out the radial mass profiles realized in simulations. But the comparisons we do make suggest that real clusters may have higher central concentrations than simulated clusters (Comerford and Natarajan 2007, Broadhurst et al. 2008, Oguri et al. 2009), perhaps indicating earlier formation times than predicted.

Recently obtained data enables us to move beyond measurements of clusters' radial profiles to studies of their substructure. Deep (20-orbit) multiband ($g'r'i'z'$) HST/ACS images of Abell 1689 reveal over 100 multiple images of strongly lensed galaxies, allowing us to produce detailed dark matter mass maps.

Halkola et al. (2007) used parameterized mass models to put constraints on subhalo truncation radii. The individual subhalos cannot be sufficiently resolved with current data, so the conclusions hinge on the accuracy of the employed models.

Natarajan et al. (2007) attempts the most direct comparison to date of the mass function in observed and simulated clusters. A broad agreement is found (see below), though both measurements have large uncertainties, and a bias of 2x must be corrected for in the measurements from simulations. The observational measurements involve strong lensing by individual galaxy halos in cluster cores while the simulation measurements involve friend-of-friend detection of substructure. The former suffers from small number statistics while the latter is plagued by the inability to cleanly extract and measure the mass of substructures embedded in the overall dark matter halo.

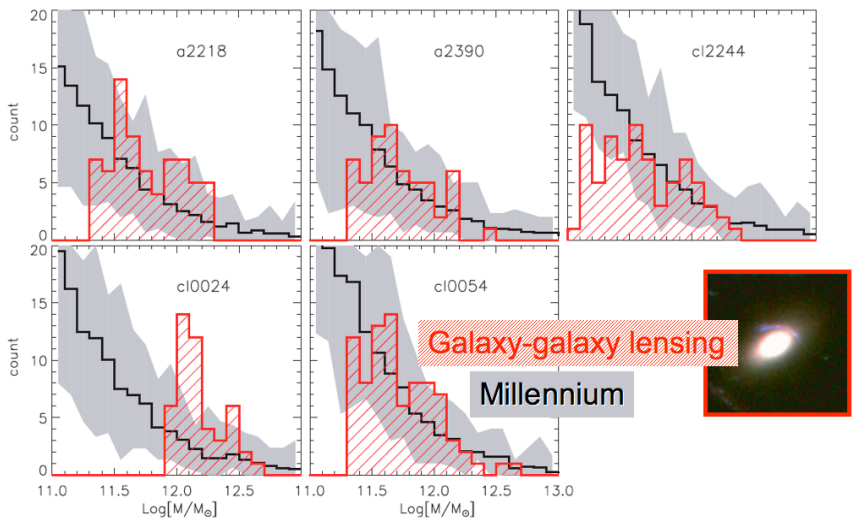

*Substructure mass function in observed and simulated galaxy clusters. There is broad agreement but with large uncertainties on both fronts. (Reprinted from Natarajan 2007 with captions and image added.)*

## Direct comparison of observed and simulated cluster substructure

We propose new direct comparisons of substructure in observed and simulated galaxy cluster cores. Gravitational lensing yields a map of total mass projected along the line of sight. Similar maps can be produced from simulated dark matter halos (see cover images). These 2-D maps can then be analyzed using the same technique and compared. Ideally the analysis will yield measures of the substructure mass power spectrum.

This is a new problem for us, as previously the data and analysis techniques have been lacking. The exact details of how to analyze such data are being considered now (Coe et al., in prep.). As of this writing, we have yet to determine exactly what parameters we will constrain of, say, the dark matter particle. Perhaps we will learn less about dark matter itself and more about the formation of galaxy clusters. The influence of baryons may also be determined, though they are normally assumed insignificant as cooling in clusters is inefficient.

Williams & Saha (2004) and Saha et al. (2007) have developed a few measures of substructure for use with their minimal-assumption mass maps. We look forward to applying these to the higher resolution mass maps to come with improved data in the next decade.

## Dark Subhalos and Other Observables

The claimed agreements between observed and simulated large scale structure (e.g., Spergel et al. 2006) are based on the assumption that light traces mass. This assumption appears to hold in general, but the degree to which light traces mass is an interesting open question. And some of the most exciting and provocative discoveries about dark matter come from weak lensing analyses in which those assumptions are dropped entirely (e.g., Clowe et al. 2006, Jee et al. 2007, Massey et al. 2007).

Our high-resolution minimal-assumption mass maps will measure the degree to which light traces mass in cluster cores. And they will enable us to search for dark subhalos devoid of luminous galaxies. Dark subhalos are theorized (in galaxies at least – Diemand et al. 2005), but the existence of one has never been proven. And with sufficiently resolved dark matter subhalos, we will be able to measure the mass-to-light ratios of individual galaxies, and look for evidence of tidal stripping.

## Required Data

To reveal hundreds of multiple images for A1689 and other clusters, deep high-resolution multiband imaging will be required. HST/ACS is capable of obtaining images of the desired quality but for prohibitive integration times (see graph below). Larger telescopes will make these observations a reality. Based on collecting area alone, JWST will reduce the integration times by a factor of 7.

To date, deep (20-orbit) multiband HST/ACS images have been obtained for five massive ($10^{14-15}$ solar mass) clusters: A1689, A2218, A1703, CL0024, and MS1358. While A1689 reveals over 100 mutliple images, the other clusters yield on the order of 40 (Elíasdóttir et al. 2008, Richard et al. 2009, Zitrin et al. in prep., and Coe et al. in prep.). This is probably due to A1689's exceptionally high central mass concentration.

Exposure depth is the primary requirement to bring fainter lensed images into view. High resolution greatly aids in the identification of multiple images as internal structures can be resolved and matched among images. Multiple bands are important for the same reason: for colors to be matched and to obtain photometric redshifts. Spectroscopic redshifts are ideal but even a few allow proper normalization of the mass model.

The importance of high-resolution images is made apparent in recent analyses of Abell 1703 from the ground (with Subaru; Oguri et al. 2009) and space (with ACS; Richard et al. 2009). The former identified 21 multiple images while the latter revealed 53.

Adaptive optics from the ground (on TMT, for example) may be sufficient if high resolution can be maintained over the entire strong lensing region (1-2 arcmin across). Presumably a single guide star would be insufficient.

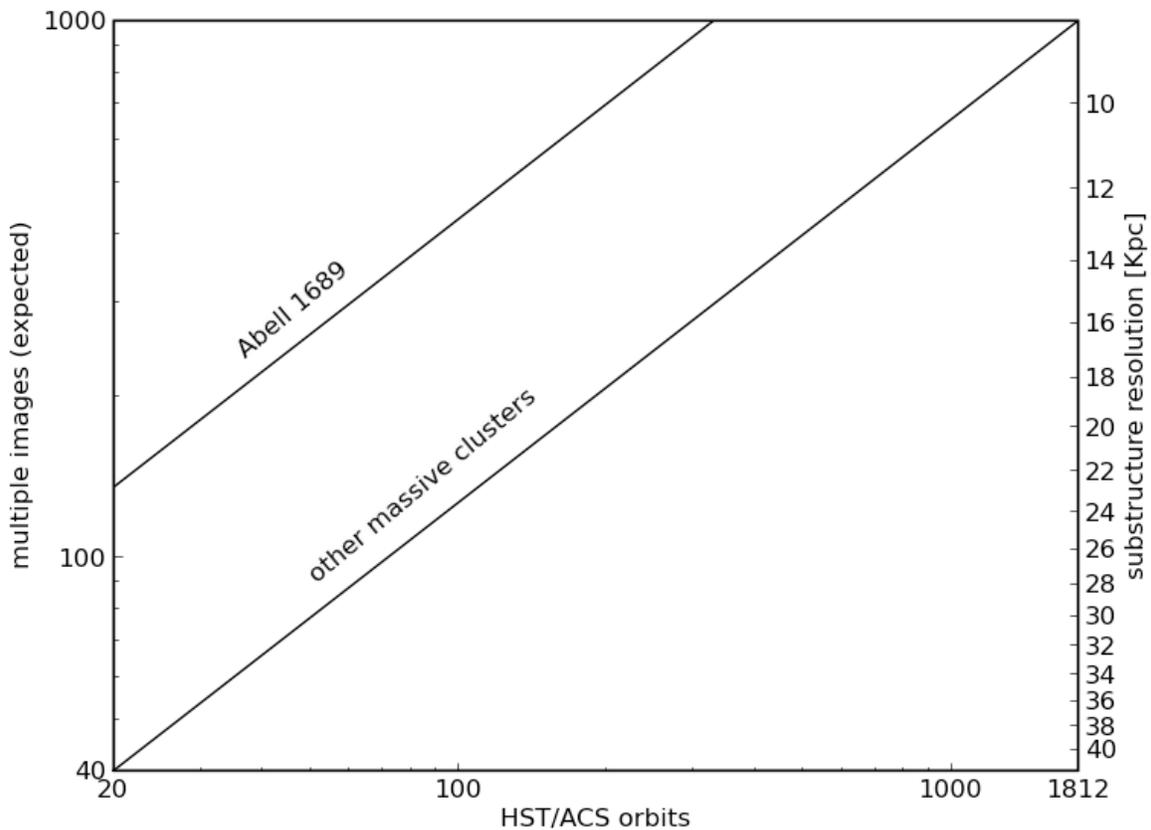

*Expected numbers of multiple images and corresponding mass resolution for deep HST/ACS observations. These integration times are prohibitive for HST but feasible with larger telescopes to come in the next decade. Abell 1689 reveals many more multiple images than other clusters of similar mass. Mass resolution is estimated as the average spacing between multiple images (see text below). Expectations were derived from extrapolations of current observations. A faint end LF slope of $\alpha = -1.7$ was assumed.*

# Analysis Method

The mass map reconstructions discussed in this paper will be made possible by a new analysis technique (LensPerfect: Coe et al. 2008). On the front cover, a mass reconstruction of Abell 1689 is presented which perfectly reproduces the observed positions of 168 knots within 135 multiple images. Only the most basic assumptions are made about the distribution of mass. Specifically, no assumptions are made about light tracing mass. The number of input multiple images is practically limitless, as an efficient direct inversion technique is employed.

This technique was only made possible recently thanks to the development of a new mathematical technique involving the interpolation of vector fields (Fuselier 2006, 2007). Each multiple image defines the deflection due to lensing at its location. We interpolate among these vectors to obtain the deflection field across the entire strong lensing region. The mass distribution is simply half the divergence of this vector field.

From this analysis, we clearly see that our mass map resolution is determined by the density of multiple images. Our linear spatial resolution is approximately equal to the average spacing between images. This is the measure given in the figure above.

Traditional strong lens mass modeling relies on assumptions of mass profile forms and light tracing mass (e.g., Limousin et al. 2007). These methods are most attractive when few constraints are available (although see Saha et al. 2006b). But as we approach the era of 1,000 multiple image systems, more flexible techniques free of the traditional assumptions are being applied to and developed for cluster strong lensing analysis (Diego et al. 2005, Saha et al. 2006, Liesenborgs et al. 2007, Coe et al. 2008).

# Summary

The next decade will witness a dramatic improvement in the quality of cluster dark matter maps. Details will be resolved on the ~10 kpc level or better, allowing us to map individual galaxy subhalos. We will verify the level to which light traces mass, search for dark subhalos devoid of luminous galaxies, measure mass-to-light ratios of individual galaxies, and look for evidence of tidal stripping. The mass power function of substructure in cluster cores will be observationally measured and compared to that found in simulations. This will provide important tests for $\Lambda$CDM on cluster scales.

# Selected References

Coe et al. (2008), ApJ, 681, 814 *LensPerfect: Gravitational Lens Mass Map Reconstructions Yielding Exact Reproduction of All Multiple Images*

Natarajan et al. (2007), MNRAS, 376, 180 *Substructure in lensing clusters and simulations*

Saha et al. (2007), ApJ 663, 29 *Meso-Structure in Three Strong-Lensing Systems*

Halkola et al. (2007), ApJ, 656, 739, *The Sizes of Galaxy Halos in Galaxy Cluster Abell 1689*

Oguri et al. (2009), arXiv/0901.4372, *Subaru Weak Lensing Measurements of Four Strong Lensing Clusters: Are Lensing Clusters Over-Concentrated?*

Broadhurst & Barkana (2008), MNRAS, 390, 1647 *Large Einstein radii: a problem for ΛCDM*

Navarro et al. (2008), arXiv/0810.1522, (MNRAS submitted) *The Diversity and Similarity of Cold Dark Matter Halos*